# Magnetic Properties and Spin-orbit Coupling induced Semiconductivity in LK-99

Hua Bai[1]*, Lei Gao[1], Jianrong Ye[1], Chunhua Zeng[1]*, Wuming Liu[2]*

[1]*Institute of Physical and Engineering Science/Faculty of Science, Kunming University of Science and Technology, Kunming 650500, China*
[2]*Beijing National Laboratory for Condensed Matter Physics, Institute of Physics, Chinese Academy of Sciences, Beijing 100190, China*
*Corresponding authors: huabai@kust.edu.cn (H.Bai); chzeng83@kust.edu.cn (C.Zeng); wliu@iphy.ac.cn (W.Liu)

Recent reports of a possible room-temperature superconductor called LK-99 have generated a lot of attention worldwide. In just a few days, a large amount of experimental works attempted to reproduce this sample and verify its properties. At the same time a large amount of theoretical works have also been reported. However, many experiments have drawn different conclusions, and many theoretical results are not consistent with experimental results. For one of the structures of LK-99 with the chemical formula as $Pb_9Cu(PO_4)_6O$, many first-principles calculations did not consider spin-orbit coupling and concluded that it is a flat band metal. However, spin-orbit coupling is often not negligible in systems with heavy elements, and LK-99 contains a large amount of heavy element Pb. We performed calculations of electronic structure of $Pb_9Cu(PO_4)_6O$ with spin-orbit coupling, and the results show that it's not a metal but a semiconductor. This is consistent with many experimental results. In the ferromagnetic state $Pb_9Cu(PO_4)_6O$ is an indirect-bandgap semiconductor with a bandgap of 292 meV. Moreover, its conduction band is a flat band. At an electron doping level of 0.5 e/unit cell, $Pb_9Cu(PO_4)_6O$ becomes metallic and has a flat band with a width of only 25 meV at the Fermi level in the ferromagnetic state. While in the antiferromagnetic-A state, $Pb_9Cu(PO_4)_6O$ is a direct-bandgap semiconductor with a bandgap of 300 meV. As a magnetic narrowband semiconductor，$Pb_9Cu(PO_4)_6O$ may have potential application value in the field of optoelectronic device, photocatalytic, photodetector and spintronics device.

**Keywords**： Superconductor, LK-99, First-principles calculation, Flat band, Spin-orbit coupling

## I. INTRODUCTION

Recently, a group from South Korea reported a possible room-temperature and atmospheric-pressure superconductor LK-99 in the form of a preprint, which has attracted great attention worldwide[1]. Superconductors have always been very popular research objects due to their zero resistance and complete diamagnetism (Meissner effect). Since Onnes discovered superconductivity in Hg in 1911 with a critical temperature ($T_C$) of 4.2 K[2], the search for room-temperature, atmospheric-pressure superconductors has been ongoing ever since. Since then, many high-temperature superconducting systems have been discovered, such as cuprate superconductors[3], iron-based superconductors[4], high-pressure hydride superconductors[5]. However, superconductors at room temperature and atmospheric pressure have never been discovered. If the superconductivity of LK-99 is confirmed, it will be an amazing discovery.

As reported, LK-99 is a compound that replaces part of Pb with Cu in lead apatite, and its chemical formula is $Pb_{10-x}Cu_x(PO_4)_6O$ (0.9 < x < 1.1)[1]. After this report, large amount of experimental works were carried out on a global scale to try to reproduce LK-99 and study its properties[6-10]. However, different experimental groups have drawn many different conclusions, making the nature of LK-99 confusing. For example, Chang et al. successfully grew LK-99 and observed diamagnetism transition and half levitation under magnetic field[10]. Jia et al. also observed half levitation but measured ferromagnetism[6]. Awana et al. also measured diamagnetism but no superconducting signal at all[8], while Shi et al. observed nearly zero resistance above 100 K but no diamagnetism[7]. Furthermore, in samples containing $Cu_2S$, Luo et al. observed the reduction in resistivity caused by the first order structural phase transition of $Cu_2S$ at around 385 K[9]. At the same time, a large number of theoretical results have also been reported[11-20], and most assume that LK-99 is a metal with two flat bands at the Fermi level. However, all previous first-principles calculations have not considered the spin-orbit coupling (SOC) effect. In systems with heavy elements, the SOC effect is often relatively strong[21], and there are many heavy elements Pb in LK-99, so its SOC effect should not be ignored.

In this work, we used first-principles method to calculate in detail the crystal structure and electronic structure of LK-99 in the case of x = 1 with the chemical formula of $Pb_9Cu(PO_4)_6O$. First, we identified the most stable structure and a metastable structure. Then, for the most stable structure, we calculated its electronic structure without and with considering the SOC. If the SOC is ignored, our calculations agree with most previous first-principles calculations[11-17]. However, the calculations considering SOC show that the ground state is the antiferromagnetic-A (AFM-A) state but the energy difference between its different magnetic configurations is very small. In the ferromagnetic (FM) state, SOC will cause the band inversion of two flat bands and open a bandgap, making $Pb_9Cu(PO_4)_6O$ an indirect-bandgap semiconductor, which is consistent with some current experimental results[6,22-24]. Its magnetic moment is about 1 μB and its easy-axis is *c*-axis. In addition, a small amount of electronic doping makes $Pb_9Cu(PO_4)_6O$ metallic with a very narrow flat band at the Fermi level. While in the AFM-A state, $Pb_9Cu(PO_4)_6O$ is a direct-bandgap semiconductor.

## II. RESULTS AND DISCUSSIONS
### A. Structures of $Pb_{10}(PO_4)_6O$ and LK-99

Figures 1(a) and 1(d) show the crystal structure of $Pb_{10}(PO_4)_6O$, which is the parent compound of LK-99. It has a hexagonal structure with space group 176 ($P6_3/m$). In $Pb_{10}(PO_4)_6O$, Pb atoms have two symmetrically equivalent positions: Pb1 and Pb2, as shown in the Figure 1. A unit cell contains four Pb1 atoms and six Pb2 atoms. The O atoms also have two symmetrically equivalent positions: O1 and O2. The twenty-four O1 atoms and six P atoms form six triangular cone-shaped $PO_4$ units, while the O2 atoms form a one-dimensional chain along the *c*-direction. It should be noted here that O2 atoms are all 1/4 occupied in the four fully equivalent positions. Previous first-principles calculation work has shown that $Pb_{10}(PO_4)_6O$ is a semiconductor[12], and previous experimental work has shown that the chemical formula of LK-99 is $Pb_{10-x}Cu_x(PO_4)_6O$ (0.9 < x < 1.1). In addition, previous theoretical work has demonstrated that the structure in which Pb1 atoms are occupied is more stable than the structure in which Pb2 atoms are occupied[13]. For convenience, we consider the case of x = 1. Since the positions of the four Pb1 atoms are completely equivalent, we first replaced one of them with Cu atoms. At this time, due to the change in symmetry, the positions of the four O2 atoms become unequal. In order to find the most stable O2 atom position, we choose one position to place an O2 atom each time, remove the other three O2 atoms, and then optimize the structure. In the end we only get two structures. If we put an O2 atom in position 1 or 2 as shown in Figure 1(d), it will become the same structure after optimization, as shown in Figure 1(e). Similarly, placing an O2 atom at the position 3 or 4 and then optimizing it also results in another same structure, as shown in Figure 1(f). This result shows that, in $Pb_9Cu_1(PO_4)_6O$, only two O2 atom sites are stable. In order to distinguish the two structures, we named the former as LK-99-1 and the latter as LK-99-2. Both LK-99-1 and LK-99-2 have a trigonal crystal structure with space group 143 (*P*3). In addition, compared with the parent compound, in LK-99-1 and LK-99-2, the position of Pb atoms and the $PO_4$ units away from Cu atom does not change much, but the three $PO_4$ units near the Cu atom all undergo obvious rotation. Furthermore, the total energy of LK-99-2 is lower than that of LK-99-1 in the FM and NM states by 0.26 eV and 0.21 eV, respectively. That is, LK-99-2 is the most stable structure, while LK-99-1 is a metastable structure.

## B. Electronic Structure without SOC

Some previous works have calculated the electronic structure of LK-99 without considering the SOC, and found two flat bands at the Fermi level[11-13]. To carefully verify these results, we first performed electronic state calculations without considering the SOC. Figure 2 shows the band structures of LK-99-1 and LK-99-2 without SOC for both ferromagnetic (FM) and non-magnetic (NM) states, and the Hubbard interaction U = 4 eV. The total energy of the FM state is 0.15 and 0.20 eV lower than that of the NM state in LK-99-1 and LK-99-2, respectively. After Cu substitution, both LK-99-1 and LK-99-2 become metals in the FM and the NM states. What's more interesting is that there are two flat bands near the Fermi level of these four band structures. We call the upper and lower flat bands as flat band1 and flat band2 respectively, as shown in Figure 2(c). Now that LK-99-2 is more stable, we will focus on it next. The flat band1 and flat band2 have some different characteristics: the highest energy point and the lowest energy point of flat band1 are at H point and Γ point, respectively, while those of flat band2 are at A point and the middle of K-H line. In the FM state, both flat bands are partially filled and fully spin polarized, while in the NM state, flat band1 is partially filled but flat band2 is fully occupied. In order to analyze the composition of states near the Fermi level, the corresponding projected density of states (PDOSs) are also calculated, as shown in Figure S1. In the NM state of LK-99-1 and LK-99-2, the states near the Fermi level are mainly contributed by Cu. In addition, O2, O1 around Cu and Pb2 also have partial contribution. While in the FM state of both, the states near the Fermi level are also mainly contributed by Cu, moreover, O1 and Pb2 also have partial contribution. Overall, spin polarization causes a slight change in the composition of states near the Fermi level. The above results are consistent with the previous literature reports[11-13]. However, it should be noted that in many calculations, the value of U has a great influence on the results. The value of U used in some calculations even requires experimental results to be carefully tested. Therefore, we also conducted tests with different U values, where the U values range from 0 to 3 eV. The results of these calculations are presented in Figure S2 and Figure S3. In the FM state and the NM state, the band structure and flat band width change little with the U value. The width of the flat bands in all cases is summarized in Figure S6(a). The width of these flat bands of LK-99-2 is roughly between 50 meV and 160 meV. Some regularities can also be drawn from these results without SOC. First, the width of flat band2 is narrower than flat band1 in all cases. Second, spin polarization will reduce the width of the flat bands. Third, as the U value increases, the width of the flat bands will also increase slightly. In general, if the SOC is not considered, the flat bands in LK-99-2 are very robust, and the conclusions are not affected by the calculation parameters.

### C. Electronic Structure with SOC

The calculation results without considering the SOC look very interesting. But it should be noted that Pb is a heavy element with a relative atomic mass of 207.2, so the effect of SOC is often not negligible[21]. It is necessary to consider the electronic structure calculation with SOC. The direction of the magnetic moment is very important in calculations of the FM state with SOC, so the magnetic anisotropy energy (MAE) is calculated first. The MAE is defined as $E_{hard} - E_{easy}$, while $E_{hard}$ and $E_{easy}$ are the total energy of the system when the magnetic moment is parallel to the hard-axis and easy-axis, respectively. Calculations prove that the easy-axis of LK-99-2 is *c*-axis while hard-axis in the *ab*-plane and perpendicular to *b*-axis. The MAE is calculated as 1.03 meV. Cu atoms contribute most of the magnetic moment. The three nearest neighbor O1 atoms around the Cu atoms also contribute a small amount of magnetic moment. The total magnetic moment is about 1 μB/unit cell. Other common antiferromagnetic (AFM) configurations are also considered, as shown in Figure S4. The total energy difference of all magnetic configurations is very small, which may be due to the relatively large distance between Cu atoms and the relatively weak superexchange interaction. Considering that many experimental works have reported the FM state of LK-99[6,22-24], but no AFM state, we first calculated the electronic structure of the FM state. Figure 3(c) shows the band structure of LK-99-2 with SOC and the Hubbard interaction U = 4 eV in the FM state. Unlike previous calculations which ignored the SOC, the band structure shows that LK-99-2 is an indirect-bandgap semiconductor with a bandgap of 292 meV. The conduction band minimum (CBM) and the valence band maximum (VBM) are located at the middle of K-H line and L point respectively. It should be noted that the conduction band is also a flat band, with a very narrow width of 52 meV. The corresponding PDOSs are shown in Figure 3(d). This flat band is mainly contributed by Cu atoms, while O atoms also have a small amount of contribution. The highest energy point and lowest energy point of this flat band are at A point and the middle of K-H line, respectively. In addition, the shape of this flat band is very similar to the shape of flat band2 in the previous calculation without SOC. Therefore, it can be judged that this flat band is the previous flat band2. Similarly, different values of U from 0 to 3 eV were also tested as shown in Figure S5. The results show that, after considering the SOC, the value of U has a great influence on the band structure. When U = 0 eV, the band structure is very similar to the case without SOC. The situation gradually changes as the U value increases. When U = 1 eV, the position of flat band2 was raised, while the position of flat band1 was lowered, which eventually caused the band inversion of the two flat bands. At the same time, a bandgap of 0.114 eV is opened. As the U value increases, the value of the bandgap also increases, and the width of flat band2 decreases slightly, but the width of flat band1 increases rapidly and then becomes a normal band. In the end, only one flat band is left. The bandgap of LK-99-2 under different conditions are summarized in Figure S6(b). Next, the electronic structure of the AFM state with the lowest total energy, that is, the AFM-A state, is also calculated. Figure 4 shows the band structure and corresponding PDOSs of LK-99-2 in the AFM-A state. In this state, LK-99-2 is a direct-bandgap semiconductor with a bandgap of about 300 meV. Its CBM and VBM are both at K point. Its conduction band is composed of two flat bands, and its composition is basically the same as the flat band of the FM state. In the experiments, the AFM state has not been observed. This may be because the energy difference between the AFM state and the FM state is too small, and the superexchange effect is too weak, so that the direction of the magnetic moment is easily reversed by the magnetic field. Finally, the ferromagnetic state is obtained in the magnetic field. We also consider the NM state with SOC. Although we try to fix the magnetic moment to 0 in VASP, the final solution still has some magnetic moments, and a strictly NM state is not obtained. Next, the electronic structure of LK-99-1 with SOC and U = 4 eV was also calculated, as shown in Figures 3 (a) and (b). Similar to LK-99-2, after considering SOC, LK-99-1 is also an indirect bandgap semiconductor with a bandgap of 190 meV in FM state.

### D. Discussion

The flat bands near the Fermi level are considered to be one of the reasons for high-temperature superconductivity[25], because these flat bands will cause a large DOS near the Fermi level. According to the Bardeen-Cooper-Schrieffer (BCS) theory, this is beneficial to increase the $T_C$ of superconductivity[26]. Flat bands are essentially the localization of electrons in real space. And there are several different mechanisms of flat band in different systems, such as heavy-fermion systems[27], twisted two-dimensional systems[28], kagome lattice systems[29] and doped semiconductors or insulators[30]. If SOC is not considered, our calculation results and some previous calculation work are more in line with this situation, i. e., there are flat bands at the Fermi level. And this flat band is more likely to be caused by doping in the insulator, because the largest contribution to this flat band is Cu, that is, most of the states near the Fermi level are localized on Cu atoms. But as mentioned above, in a system with 9 Pb atoms in a unit cell, ignoring the SOC is imprecise and the results are often unreliable. In fact, some experiments trying to reproduce LK-99, the result of the resistance measurement is a semiconductor[6,9,23,31], which is not consistent with the calculated metallic behavior. In addition, as a flat band system, LK-99-2 should have a relatively strong electron correlation, so we are more inclined to the calculation results with a relatively large U value, i.e, LK-99-2 is a semiconductor both in the FM state and in the AFM-A state. As a magnetic narrowband semiconductor, LK-99-2 also has some possible potential applications. For example, optoelectronic device[32], photocatalytic[33], photodetector[34], spintronics device[35]. Additionally, in order to get the flat band on the Fermi level, we can make some modulations to LK-99-2. For example, a small amount of electronic doping allows the conduction band to be partially occupied, and a flat band at the Fermi level can be obtained. Therefore, we tried to calculate the electronic doping of LK-99-2 after considering the SOC in the FM state. As shown in Figure 5, after doping 0.5 electron per unit cell, the flat band becomes partially occupied, and the width of this flat band is only 25 meV.

It should be noted that our calculations are only focus on two simple structures (LK-99-1 and LK-99-2) with the chemical formula of $Pb_9Cu_1(PO_4)_6O$, which are the second and most stable structures considering only x = 1 with the Cu atoms occupying the Pb1 sites. These are just some of the possible structures of LK-99. Depending on the number and position of Cu atoms and O2 atoms, LK-99 has other more complex structures with larger unit cells. In addition, it is impossible to judge whether a compound is a superconductor or not only from the calculations of the electronic structure. Even if there is a flat band at the Fermi level, it does not mean that it is a superconductor.

## III. SUMMARY

In conclusion, our first-principles calculations show that SOC is non-negligible in $Pb_9Cu(PO_4)_6O$. If the SOC is not considered, although the calculation can get the flat bands on the Fermi level, this result is not rigorous and does not agree with some experiments. Our calculations show that the ground state is the AFM-A state but the energy difference between different magnetic configurations is very small. Once the SOC is considered, a bandgap is opened, making $Pb_9Cu(PO_4)_6O$ an indirect-bandgap semiconductor with a bandgap of 292 meV and a flat conduction band in the FM state. After electron doping at the level of 0.5 e/unit cell, it becomes metallic and has a flat band with a width of only 25 meV at the Fermi level in the FM state. While in AFM-A state, $Pb_9Cu(PO_4)_6O$ is a direct-bandgap semiconductor with a bandgap of 300 meV. Our calculation results explain the inconsistencies of some previous experimental and theoretical works[6,9,23,31], make up for the lack of previous theoretical works[11-17], point out the importance of considering SOC in the theoretical work of LK-99, and also provide some suggestions for experimental works. In addition, our work also provides some theoretical basis for the possible potential application of $Pb_9Cu(PO_4)_6O$ as a magnetic narrowband semiconductor.

## Methods

In this work, the calculations were performed using the Vienna ab initio simulation package (VASP)[36] with the projector-augmented wave (PAW) method[37]. The generalized gradient approximation (GGA) with Perdew, Burke, and Ernzerhof (PBE)[38] realization was used for the exchange-correlation function. The cutoff energy was set above 450 eV. The force and energy convergence criteria were set to 0.01 eV/Å and $10^{-5}$ eV, respectively. A 5 × 5 × 7 Γ-centered *k*-point mesh was used for the Brillouin zone sampling. In order to better compare with the experimental results, in the process of structure optimization, we fixed the lattice constants as the experimental value: *a* = 9.843 Å and *c* = 7.428 Å[1]. The electronic correlation effects of Cu 3d orbits were described using the DFT+U method[39], and the values of U are carefully tested from 0 to 4 eV.

## Acknowledgments

This work was supported by the National Natural Science Foundation of China (Grants No. 12265017, No. 12247205, No. 61835013, No. 12174461, No. 12234012 and 12334012), National Key R&D Program of China (Grants No. 2021YFA1400900, No. 2021YFA0718300, No. 2021YFA1402100), Yunnan Fundamental Research Projects (Grants No. 202301AT070158、No. 202201AV070003 and No. 202101AS070018), Yunnan Province Ten Thousand Talents Plan Young and Elite Talents Project and Yunnan Province Computational Physics and Applied Science and Technology Innovation Team and Space Application System of China Manned Space Program.

## Conflict of interest

*The authors declare no conflict of interet*

## Author contributions

W.Liu and C.Zeng conceived and supervised the project, H.Bai and L.Gao performed the first-principles calculations, J.Ye processed the data and analyzed the literature, H.Bai wrote the manuscript. All contributed to the discussion of the data and revision of the manuscript.

**Data availability**

The data that support the findings of this study are available from the corresponding author upon reasonable request.

## Referees:

**Figures:**

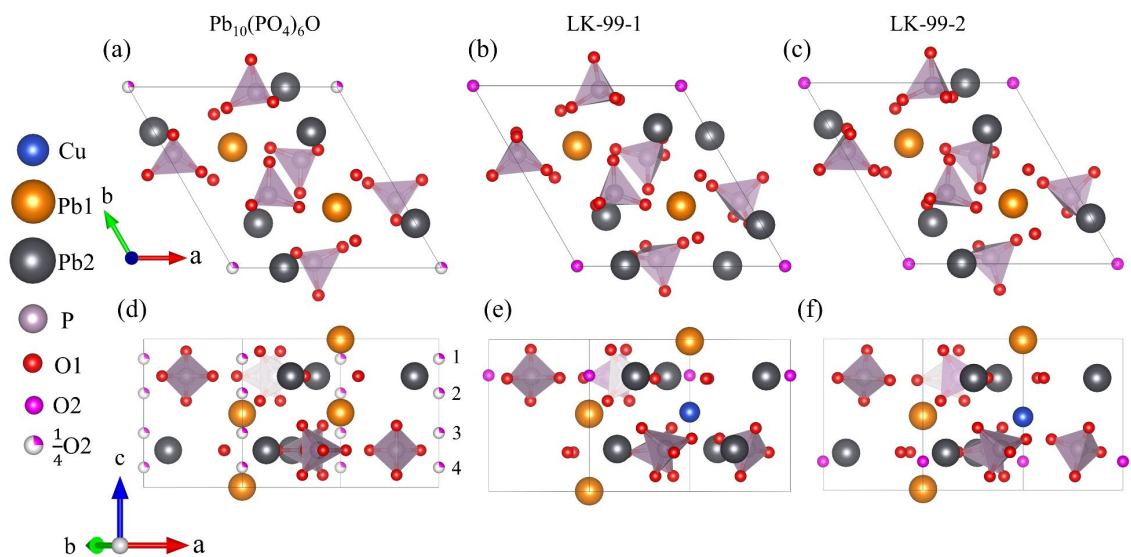

**Figure 1.** Crystal structures of $Pb_{10}(PO_4)_6O$、LK-99-1 and LK-99-2. (a) - (c) Top view. (d) - (f) Side view. In the structure of $Pb_{10}(PO_4)_6O$, the O2 atoms are 1/4 occupied in the 4 positions shown in (d).

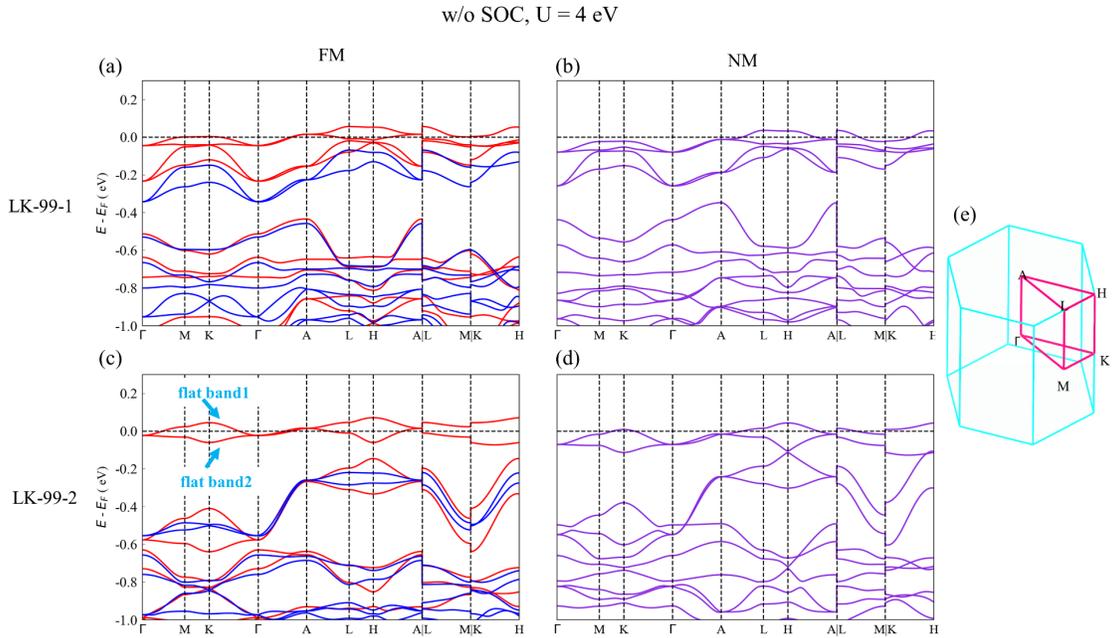

**Figure 2.** (a) – (d) Band structures of LK-99-1 and LK-99-2 without SOC (w/o SOC), and the Hubbard interaction U = 4 eV. (a) and (c) are ferromagnetic (FM) states, where the red and dark blue lines represent spin-up bands and spin-down bands, respectively. (b) and (d) are non-magnetic (NM) states. The blue arrows in (c) point to flat band1 and flat band2. (e) Brillouin zone and high symmetry points, and paths for band calculations.

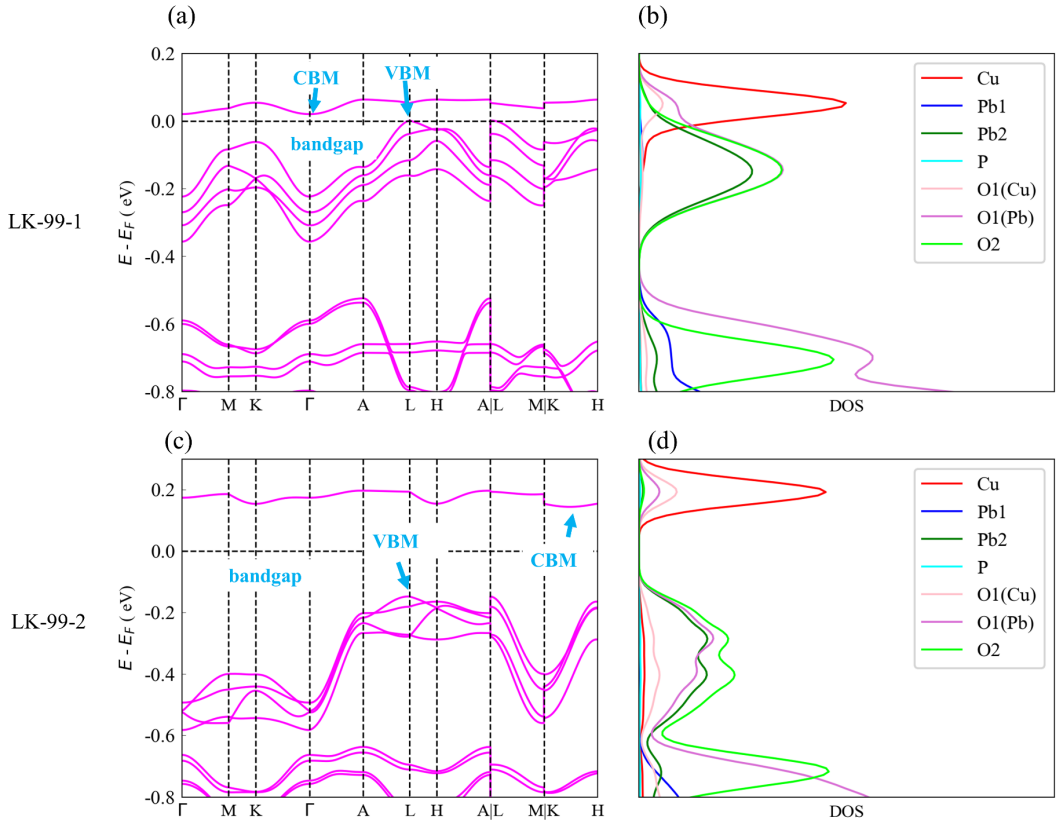

**Figure 3.** (a) (c)Band structures of LK-99-1 and LK-99-2 with SOC and the Hubbard interaction U = 4 eV in FM state. (b) (d) The corresponding projected density of states (PDOSs), O1(Cu) and O1(Pb) represent O1 atoms near Cu and near Pb, respectively.

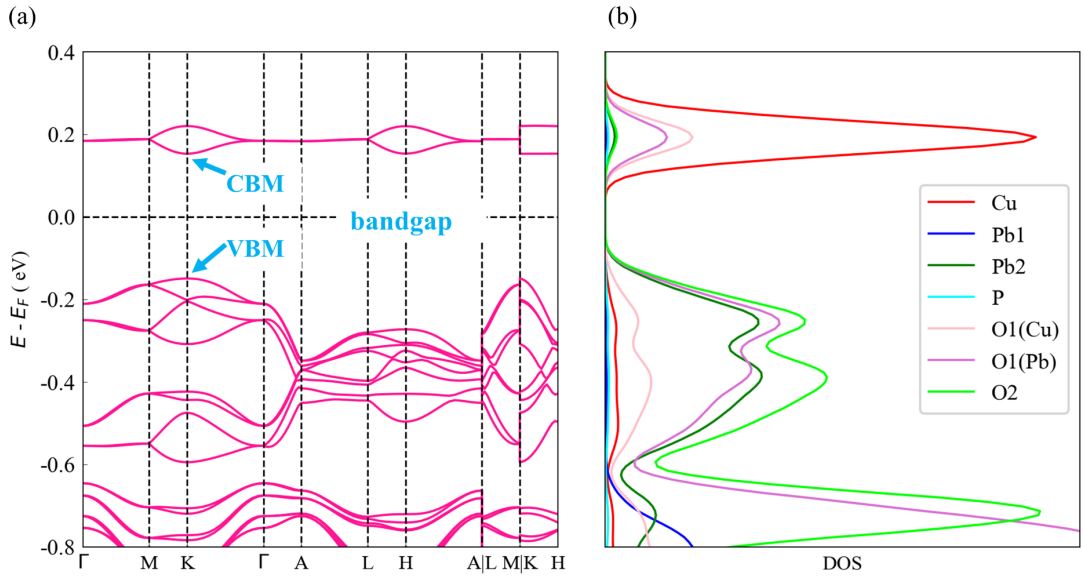

**Figure 4.** (a) Band structures of LK-99-2 with SOC and the Hubbard interaction U = 4 eV in AFM-A state. (b) The corresponding PDOSs.

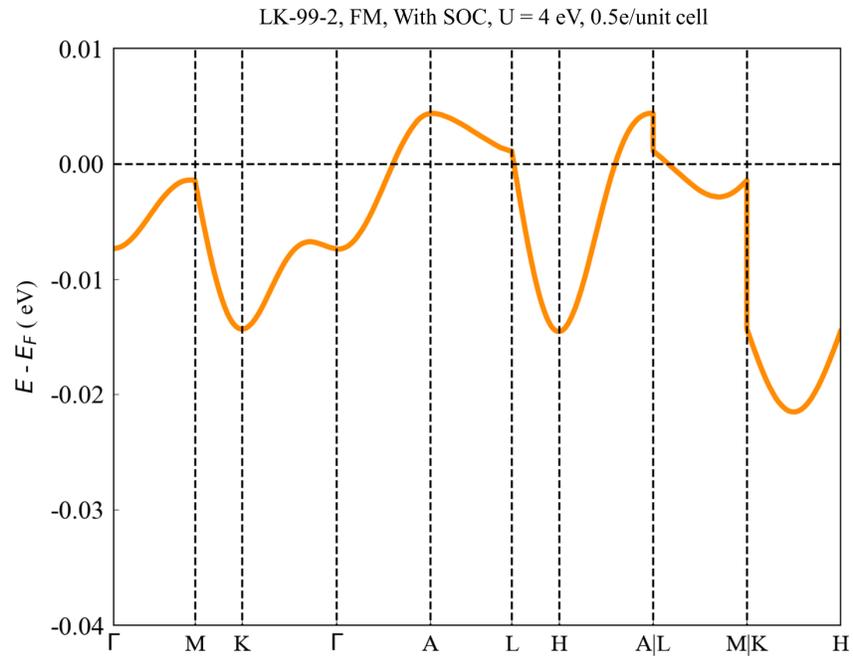

**Figure 5.** (a) The flat band on the Fermi level of LK-99-2 with SOC and the doping with 0.5 electrons per unit cell in FM state. The Hubbard interaction U = 4 eV.

# Supplementary Information
# Magnetic Properties and Spin-orbit Coupling induced Semiconductivity in LK-99

Hua Bai[1]*, Lei Gao[1], Jianrong Ye[1], Chunhua Zeng[1]*, Wuming Liu[2]*

[1]*Institute of Physical and Engineering Science/Faculty of Science, Kunming University of Science and Technology, Kunming 650500, China*
[2]*Beijing National Laboratory for Condensed Matter Physics, Institute of Physics, Chinese Academy of Sciences, Beijing 100190, China*

*Corresponding authors: huabai@kust.edu.cn (H.Bai); chzeng83@kust.edu.cn (C.Zeng); wliu@iphy.ac.cn (W.Liu)

## CONTENTS

# S1. Projected density of states (PDOS) of LK-99-1 and LK-99-2 with out (w/o) Spin-orbit Coupling (SOC)

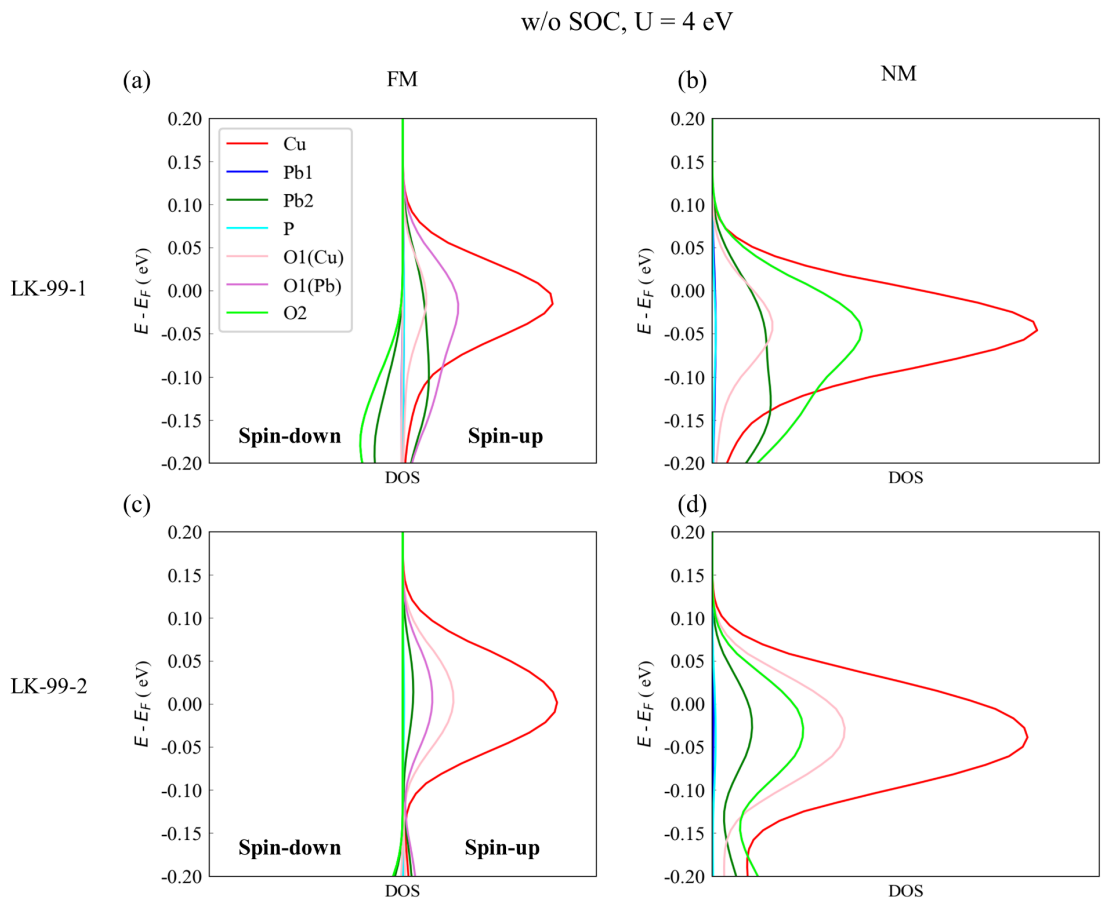

**Figure S1.** (a) – (d) Projected density of states (PDOSs) of LK-99-1 and LK-99-2 with out (w/o) SOC, and the Hubbard interaction U = 4 eV. (a) and (c) are ferromagnetic (FM) states. (b) and (d) are non-magnetic (NM) states.

## S2. Band structures of LK-99-2 w/o SOC with different U

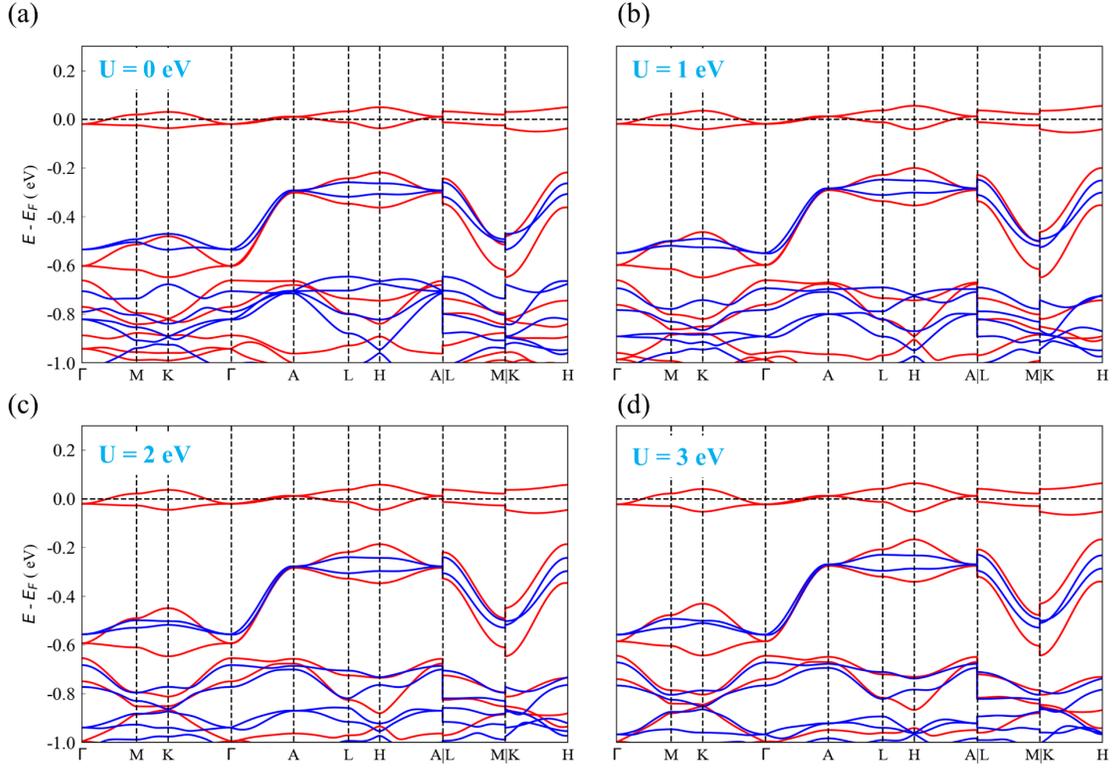

**Figure S2.** Band structures of LK-99-2 without SOC, and the different Hubbard interaction U from 0 to 3 eV in the FM state.

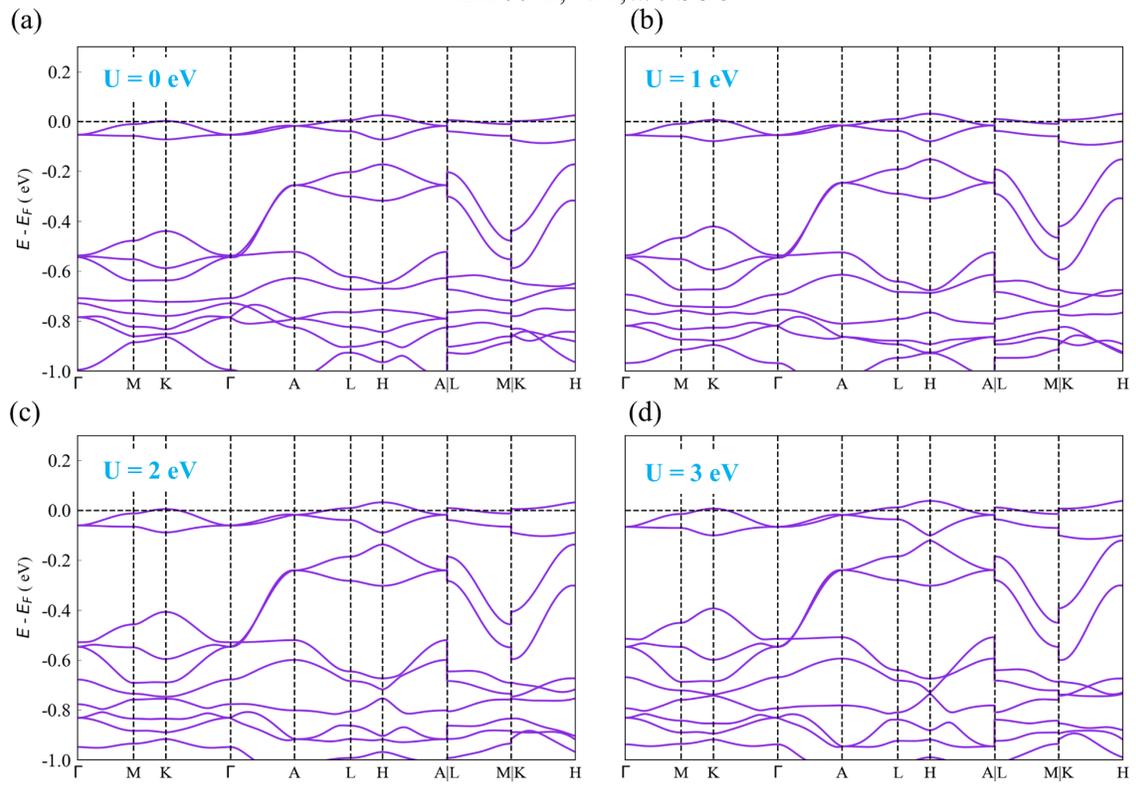

**Figure S3.** Band structures of LK-99-2 without SOC, and the different Hubbard interaction U from 0 to 3 eV in the NM state.

## S3. Schematic diagrams of different magnetic configurations

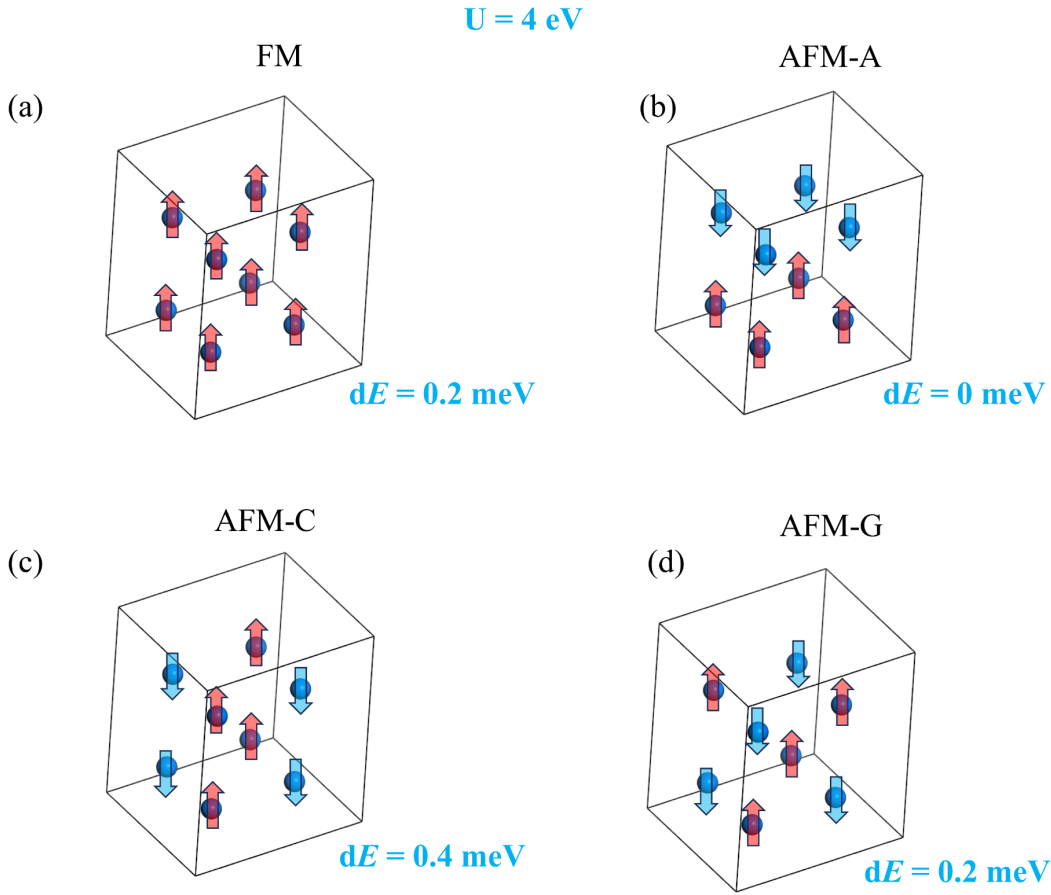

**Figure S4.** Schematic diagrams of different magnetic configurations in 2*2*2 supercell. For clarity, only Cu atoms are shown. Arrows indicate the direction of the magnetic moments. (a) Ferromagnetic (FM). (b) Antiferromagnetic-A (AFM-A). (c) AFM-C. (d) AFM-G. The blue numbers represent the relative total energys (d$E$) with Hubbard interaction U = 4 eV while the total energy of AFM-A is set to 0 meV.

## S4. Band structures of LK-99-2 with SOC with different U

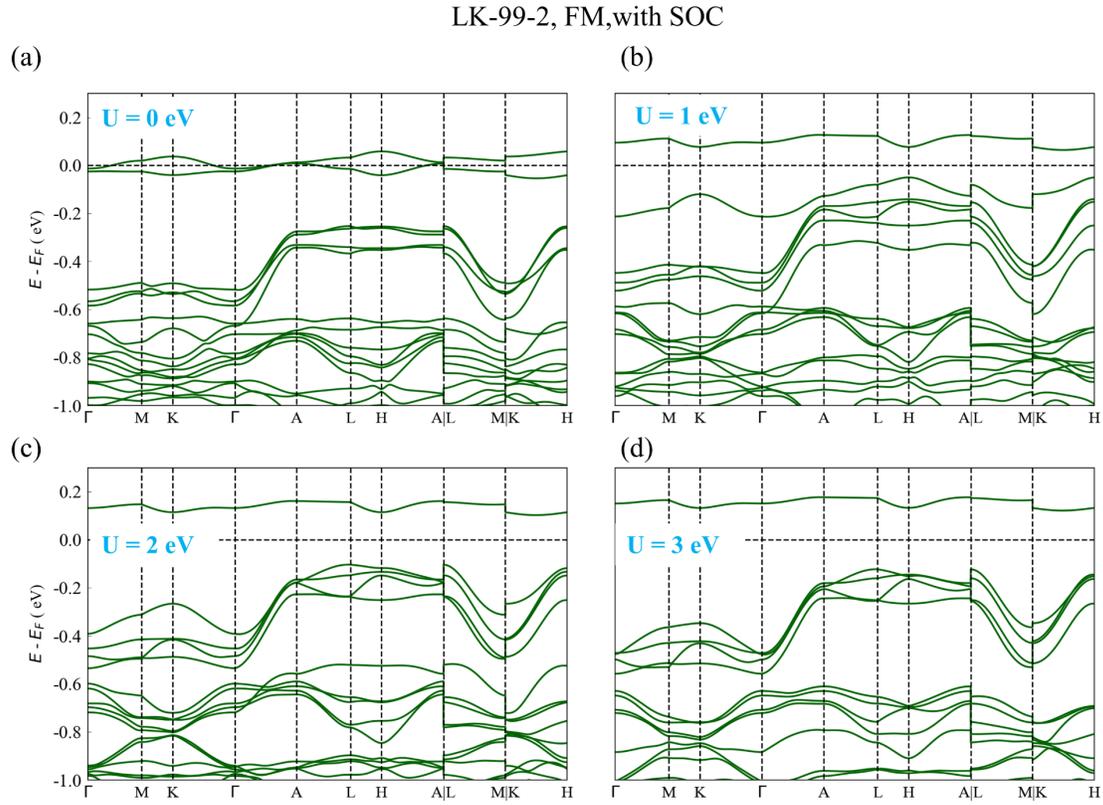

**Figure S5.** Band structures of LK-99-2 with SOC and the different Hubbard interaction U from 0 to 3 eV in the FM state. When U = 0 eV, the system is metallic. As U increases, the system becomes semiconducting and the bandgap increases.

## S5. The summary of flat band width and bandgap of LK-99-2

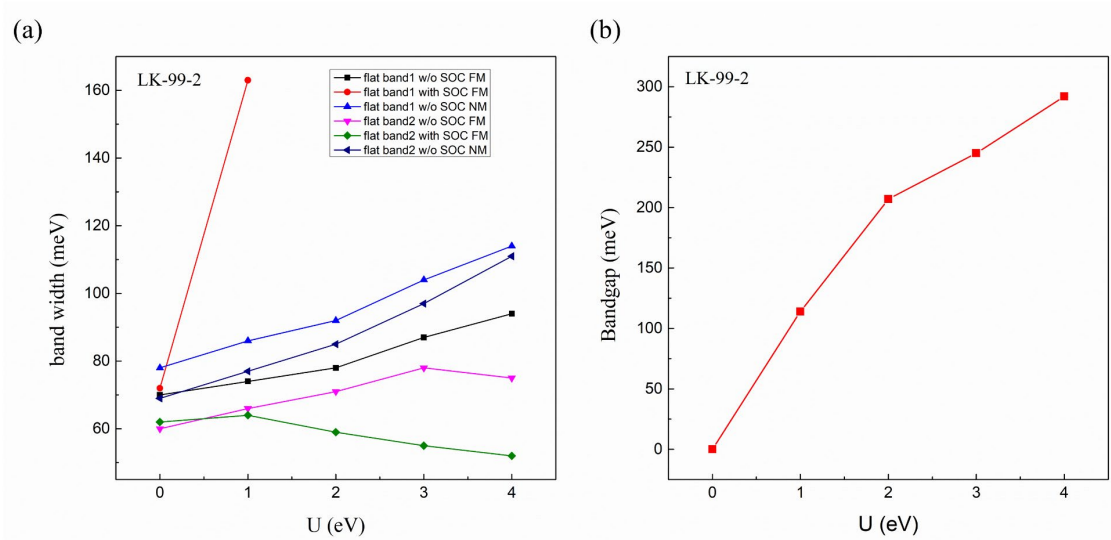

**Figure S6.** (a) The width of flat band1 and flat band2 of LK-99-2 with different Hubbard interaction U under different conditions ( w/o SOC in NM state, w/o SOC in FM state, with SOC in FM state). FM and NM represent the ferromagnetic and non-magnetic states, respectively. (b) The summary of bandgap of LK-99-2 under different U from 0 to 4 eV in FM state. The bandgap increases with the increased value of U.